\documentclass[conference]{IEEEtran}
\IEEEoverridecommandlockouts
\usepackage{cite}
\usepackage{amsmath,amssymb,amsfonts}
\usepackage{algorithm}
\usepackage{algpseudocode}
\usepackage{graphicx}
\usepackage{textcomp}
\usepackage{xcolor}
\usepackage[a4paper, total={184mm,239mm}]{geometry}
\def\BibTeX{{\rm B\kern-.05em{\sc i\kern-.025em b}\kern-.08em
    T\kern-.1667em\lower.7ex\hbox{E}\kern-.125emX}}

\usepackage{makecell}
\usepackage{multirow}
\usepackage[skip=2pt]{caption}

\usepackage[hidelinks]{hyperref}
\usepackage{booktabs}
\usepackage{pifont}
\usepackage{subfigure}
\usepackage{orcidlink}

\begin{document}

\title{SimFuzz: Similarity-guided Block-level Mutation for RISC-V Processor Fuzzing}

\author{
  \IEEEauthorblockN{
    Hao Lyu\,\orcidlink{0009-0007-0580-3346}, 
    Jingzheng Wu\textsuperscript{\href{mailto:jingzheng08@iscas.ac.cn}{\ding{41}}}\,\orcidlink{0000-0001-5561-9829}, 
    Xiang Ling\textsuperscript{\href{mailto:lingxiang@iscas.ac.cn}{\ding{41}}}\,\orcidlink{0000-0002-7377-7844}, 
    Yicheng Zhong\,\orcidlink{0009-0004-2505-3755},
    Zhiyuan Li\,\orcidlink{0009-0000-0001-7097},
    Tianyue Luo\,\orcidlink{0000-0001-7407-8255}
  }
  \IEEEauthorblockA{
    University of Chinese Academy of Sciences \\
    Institute of Software Chinese Academy of Sciences
  }
}
\maketitle

\begin{abstract}

    The Instruction Set Architecture (ISA) defines processor operations and serves as the interface between hardware and software. 
    As an open ISA, RISC-V lowers the barriers to processor design and encourages widespread adoption, but also exposes processors to security risks such as functional bugs.
    Processor fuzzing is a powerful technique for automatically detecting these bugs. However, existing fuzzing methods suffer from two main limitations. First, their emphasis on redundant test case generation causes them to overlook cross-processor corner cases. Second, they rely too heavily on coverage guidance. Current coverage metrics are biased and inefficient, and become ineffective once coverage growth plateaus.

    To overcome these limitations, we propose SimFuzz, a fuzzing framework that constructs a high-quality seed corpus from historical bug-triggering inputs and employs similarity-guided, block-level mutation to efficiently explore the processor input space. By introducing instruction similarity, SimFuzz expands the input space around seeds while preserving control-flow structure, enabling deeper exploration without relying on coverage feedback.
    We evaluate SimFuzz on three widely used open-source RISC-V processors: Rocket, BOOM, and XiangShan, and discover 17 bugs in total, including 14 previously unknown issues, 7 of which have been assigned CVE identifiers. These bugs affect the decode and memory units, cause instruction and data errors, and can lead to kernel instability or system crashes. Experimental results show that SimFuzz achieves up to 73.22\% multiplexer coverage on the high-quality seed corpus.
    Our findings highlight critical security bugs in mainstream RISC-V processors and offer actionable insights for improving functional verification.
    
\end{abstract}


\section{Introduction}
    The ISA serves as the fundamental bridge between hardware and software, defining the complete set of operations that a processor can interpret and execute~\cite{whatisisa}. As an open and freely available ISA, RISC-V lowers barriers to processor design and has driven widespread adoption, with millions of processors now deployed~\cite{riscv_milestone_2024}. At the same time, RISC-V processors face increasingly severe security challenges. 
    One example is GhostWrite~\cite{riscvuzz}, a critical bug in T-Head XuanTie~\cite{xuantie-c910} RISC-V processors. It allows attackers to bypass privilege enforcement, gain arbitrary memory access, and control peripheral devices. 
    The root cause lies in the XuanTie processor’s support for a non-standard instruction, which violates the ISA specification.
    Such implementation-specific bugs typically stem from design flaws at the processor’s Register Transfer Level (RTL) and ultimately manifest as functional bugs.
    These bugs directly affect software correctness and can occur even in the absence of adversarial inputs.
    Unlike software, processors cannot be patched once fabricated~\cite{hardfails}. Because the RTL’s functionality and timing behavior remain fixed after tape-out, comprehensive RTL-level verification is essential to ensure correctness and enhance security before fabrication.
    
    Modern processors are extremely complex, making exhaustive exploration of their state space infeasible. As a result, processor fuzzing~\cite{Kevin18rfuzz,hur2021difuzzrtl,thehuzz,processorfuzz,morfuzz,pathfuzz,fuzzhwlikesw,ssfuzz,directfuzz,cascade} has emerged as an effective and scalable method for uncovering RTL-level bugs. Current fuzzers typically rely on coverage metrics to track the proxy of the processor state. For example, multiplexer (mux) coverage~\cite{Kevin18rfuzz} focuses on multiplexer control signals. During fuzzing, these coverage metrics guide mutation, enabling exploration of a broader range of processor states through mutated seeds.

    While promising, coverage-guided processor fuzzers face two major limitations. First, by generating large numbers of redundant test cases, they often fail to capture cross-processor corner cases. For instance, PathFuzz~\cite{pathfuzz} uses real-world programs (e.g., SPEC CPU2006) to expand its seed corpus and fuzz the XiangShan~\cite{XiangShanWebsite} processor. Even with a large number of seeds, it still failed to uncover certain corner-case bugs, such as bug B7 (Table~\ref{tab:buglist}) identified in our study.
    Second, fuzzers rely too heavily on coverage guidance. ProcessorFuzzer~\cite{processorfuzz} and DifuzzRTL~\cite{hur2021difuzzrtl} design new coverage metrics to guide fuzzing, but the choice of metrics often reflects human bias, limiting verification effectiveness~\cite{coverageMetricsAnalysis}. Selecting too few metrics results in insufficient verification, while tracking too many increases overhead and reduces efficiency. Moreover, when coverage plateaus~\cite{covsaturation}, fuzzers lose meaningful guidance, severely restricting exploration. Meanwhile, increased coverage does not necessarily translate to effective bug discovery~\cite{covWithBug}.

    Because processor states are driven by inputs, broader state exploration requires a large set of high-quality test cases that can exercise diverse states. We therefore formulate processor verification as the problem of exploring the processor’s input space. To this end, we propose SimFuzz. 
    First, SimFuzz builds a high-quality seed corpus by collecting bug-triggering test cases from real-world processors, ensuring that exploration starts in regions of the input space more likely to expose bugs. Second, it preserves the control-flow structure of these seeds during mutation, maintaining semantic validity. Finally, SimFuzz introduces instruction similarity as a guidance mechanism. Unlike coverage metrics, similarity captures fine-grained variations between seeds, enabling more effective input space exploration.

    We evaluate SimFuzz on three widely used open-source RISC-V processors: Rocket~\cite{rocketrepo}, BOOM~\cite{boomrepo}, and XiangShan~\cite{XiangShanWebsite}. SimFuzz achieves 73.22\% mux coverage on the constructed seed corpus. Compared with seeds generated by Csmith~\cite{csmith-repo} and Cascade~\cite{cascade}, SimFuzz slightly outperforms PathFuzz~\cite{pathfuzz} in coverage. More importantly, SimFuzz discovered 17 bugs in total, including 14 previously unknown issues, 7 of which have been assigned CVE identifiers. These bugs affect the decode and memory units, cause misdecoding and data errors, and can lead to kernel instability or system crashes.

    In summary, this paper makes three key contributions:
    \begin{itemize}
        \item We construct a real-world, high-quality seed corpus from bug-triggering test cases across three open-source RISC-V processors: Rocket, BOOM and XiangShan.
        
        \item We present SimFuzz\footnote{https://github.com/has2lab/SimFuzz}, a new processor fuzzing framework that replaces coverage-guided guidance with similarity-guided block-level mutation.

        \item Our evaluation shows that SimFuzz achieves higher coverage and discovers 14 previously unknown bugs, 7 of which have been assigned CVE identifiers.
    \end{itemize}

\section{Background}

\subsection{RISC-V}
 \label{subsec:riscv}


    RISC-V is an open ISA designed for modularity and extensibility~\cite{manual_unprivileged, manual_privileged}. A processor can implement only the minimal base integer instruction set or extend it with additional capabilities. To ensure ecosystem consistency, RISC-V defines a standard general-purpose configuration called the G extension set, which includes Integer, Multiplication/Division, Atomic, and Floating-Point (Single and Double precision) extensions.

    The G extension specifies six instruction formats~\cite{manual_unprivileged}: Register-Register (R-type), Register-Immediate (I-type), Store (S-type), Conditional Branch (B-type), Upper Immediate (U-type), and Unconditional Jump (J-type). Table~\ref{tab:riscv_instruction_formats} summarizes these formats for 32-bit instructions.

      \begin{table}[h!]
        \centering
        \caption{RISC-V Instruction Format}
        \label{tab:riscv_instruction_formats}
        \begin{tabular}{|c|c|c|c|c|c|c|c|}
        \hline
        \text{31:25} & \text{24:20} & \text{19:15} & \text{14:12} & \text{11:7} & \text{6:0} & \text{Type} \\
        \hline
        funct7 & rs2 & rs1 & funct3 & rd & opcode & R \\
        \hline
        \multicolumn{2}{|c|}{imm} & rs1 & funct3 & rd & opcode & I \\
        \hline
        imm & rs2 & rs1 & funct3 & imm & opcode & S \\
        \hline
        imm & rs2 & rs1 & funct3 & imm & opcode & B \\
        \hline
        \multicolumn{4}{|c|}{imm} & rd & opcode & U \\
        \hline
        \multicolumn{4}{|c|}{imm} & rd & opcode & J \\
        \hline
        \end{tabular}
        \end{table}

    The semantics of the instruction determine the state of the processor. In RISC-V instructions, the \texttt{opcode} and \texttt{funct} fields jointly determine semantics. The \texttt{opcode} specifies the instruction format and basic functionality, while the \texttt{funct} field refines precise operation executed by the processor.
    In practice, the processor’s internal state space is not directly observable from the input space, making it difficult to reason about state-space coverage explicitly.
    However, processor state transitions are fundamentally driven by instruction semantics: the processor interprets each instruction according to its semantic definition and updates both its architectural and microarchitectural states.
    Therefore, even though the exact internal states are inaccessible, the semantic properties of instructions can serve as a reliable proxy for predicting their effects on processor states.


    %

\subsection{Processor Fuzzing}
    The rapid growth of open-source RISC-V processors has made processor fuzzing an increasingly important verification technique. Fuzzing systematically explores processor states by mutating seeds and observing deviations from expected behavior.
    RFUZZ~\cite{Kevin18rfuzz} was the first hardware fuzzing framework, using mux coverage to verify RTL correctness. 
    Building on this foundation, subsequent research has explored a variety of input generation and mutation techniques.

    Generative processor fuzzing requires the creation of a large number of test cases.
    Cascade~\cite{cascade} introduces an asymmetric ISA pre-simulation mechanism that intertwines control and data flows during input generation. However, its seeds often cluster close together in the input space, limiting diversity and slowing coverage growth.
    ChatFuzz~\cite{chatfuzz} and GenHuzz~\cite{genhuzz} apply reinforcement learning to generate test cases. While promising, these approaches incur high computational and time costs and struggle to explore rare but critical states. In summary, although generative fuzzing can produce many seeds and achieve high coverage, it introduces substantial redundancy and remains ineffective at exposing certain corner cases.

   Coverage-guided mutation fuzzing modifies existing seeds at the instruction level to expand the input space and uncover corner-case processor states. DifuzzRTL~\cite{hur2021difuzzrtl}, TheHuzz~\cite{thehuzz}, and MorFuzz~\cite{morfuzz} apply per-instruction mutations guided by coverage metrics. However, these techniques often lose the semantic structure of the original seeds, disrupting control flow and producing invalid or less effective test cases. PathFuzz~\cite{pathfuzz} preprocesses seeds by converting memory from a linear to a footprint layout and then uses LibAFL~\cite{libafl} to guide mutation with coverage metrics. As a software fuzzing library, LibAFL lacks hardware awareness. Overall, these coverage-guided mutation approaches rely heavily on coverage metrics and overlook the semantic structure of seeds, limiting their ability to fully exploit the information embedded in existing test cases.



\section{SimFuzz}
\begin{figure*}[ht]
      \centering
      \includegraphics[width=0.95\textwidth]{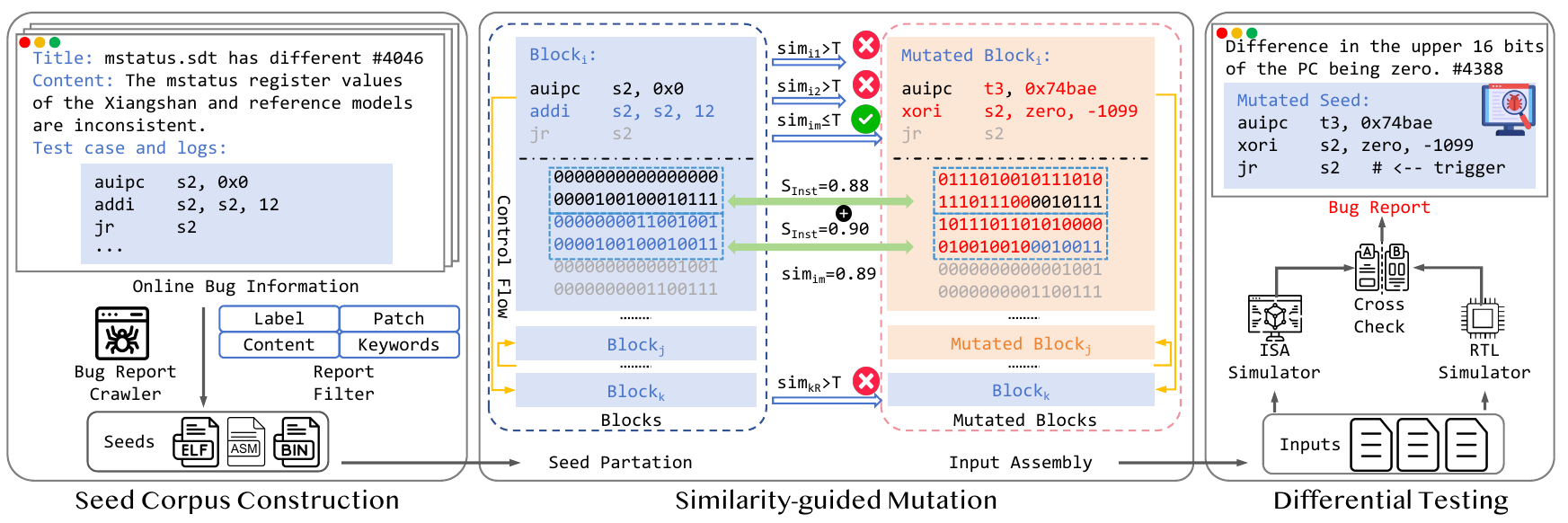}
      \caption{Overview of SimFuzz workflow.}
      \label{fig:overview}
    \end{figure*}

    SimFuzz is a processor fuzzing framework that explores the input space using a similarity-guided mutation strategy, rather than relying on coverage feedback. As shown in Fig.~\ref{fig:overview}. SimFuzz begins by constructing a high-quality seed corpus (Section~\ref{subsec:seedcorpusconstruction}) from test cases that previously triggered real bugs. During mutation (Section~\ref{subsec:mutation}), it preserves each seed’s control-flow structure while expanding the input space based on instruction similarity. Finally, the mutated seeds are evaluated through differential testing (Section~\ref{subsec:difftest}) to expose inconsistencies between the processor under test and ISA simulators.
    
\subsection{High-quality Seed Corpus Construction}
\label{subsec:seedcorpusconstruction}

    The quality of the seed corpus strongly affects fuzzing effectiveness~\cite{Klees18evaluatingFuzz}. High-quality seeds guide the fuzzer toward critical execution paths more quickly, increasing the likelihood of early bug discovery. Unlike earlier fuzzers~\cite{Kevin18rfuzz, fuzzhwlikesw, processorfuzz} that used empty corpora or focused on artificially generated seeds~\cite{cascade, genhuzz, chatfuzz}, SimFuzz builds its corpus directly from real-world bug-triggering test cases.
    The key insight is that processor bugs often arise from misinterpretations of the ISA specification or subtle design deviations. By reusing test cases that have already triggered real bugs, SimFuzz ensures that exploration starts from meaningful points in the input space rather than random points.

    To build its seed corpus, SimFuzz implements an online bug-report crawler and filter targeting three major open-source RISC-V processors: Rocket~\cite{rocketrepo}, BOOM~\cite{boomrepo}, and XiangShan~\cite{xiangshan}. Bug information is gathered from GitHub issues and CVE entries. GitHub reports are automatically filtered based on tags, patches, titles, and descriptions, then categorized by available testing resources: some contain complete executable test cases, others include partial code snippets requiring completion, and some provide only textual descriptions, necessitating manual analysis and test case construction. Similarly, CVE entries from the past decade are analyzed to identify relevant bugs and reconstruct test cases when possible.
    All resulting test cases, whether executable, completed, or manually created, are consolidated into a real-world seed corpus, which forms the foundation of SimFuzz.

    These seeds not only enable high-quality fuzzing but also highlight error-prone instructions and functional units in RISC-V processors, providing valuable guidance for both researchers and vendors.

\subsection{Similarity-guided Mutation}
\label{subsec:mutation}
    All seeds in the seed corpus are initial anchor in processor input space exploration.
    To expand the input space while preserving semantic structure of seeds, SimFuzz introduces similarity-guided block-level mutation.
    
\subsubsection{Block-level Mutation}
\label{subsub:blockmutation}

    A block is defined as a continuous sequence of instructions with a single exit point, such as jumps, branches, or system calls, which alters the control flow. The exit of a block is a Control Transfer Instruction (CTI), whose position determines the block boundary. Formally, for a given seed $S$, the seed is segmented as $N$ blocks: $S = \{B_1, B_2, ..., B_N\}$. Each block $B_i$ contains a fixed number $M_i$ of instructions ($Inst$), such that for each $i \in N$, $B_i = \{Inst_1, Inst_2, ..., Inst_{M_i-1}, CTI\}$.
    CTIs include target addresses, and modifying them can directly affect the processor’s control flow. Changes to CTIs may cause test programs to crash or deviate from the intended execution path, limiting deep exploration of the instruction stream. Therefore, block mutation deliberately avoids altering CTIs, preserving the seed’s original control flow as much as possible.

    During block mutation, SimFuzz selects several instructions within the block for individual instruction-level mutation. The mutated instructions replace their original counterparts and are reinserted into the block, producing a new, mutated version. After mutation, the similarity between the mutated block and the original block is computed based on instruction-level similarity, as defined in Algorithm~\ref{alg:block_similarity}.

         \begin{algorithm}
        \caption{Block Similarity Calculation Function}
        \label{alg:block_similarity}
        \begin{algorithmic}[1]
        
        \Function{BlockSimilarity}{$B_1$, $B_2$}
            \State $sim \leftarrow 0.0$; $size \leftarrow \min(|B_1|, |B_2|)$
            \For{$i = 1$ to $size$}
                \State $sim \leftarrow sim + \text{S}(B_1[i], B_2[i]) \times (B_1[i] \neq B_2[i])$
            \EndFor
            \State \Return ${sim}/{size}$
        \EndFunction
        
        \end{algorithmic}
        \end{algorithm}

\subsubsection{Similarity-guided}
\label{subsub:similarity}
    The core idea is to mutate blocks while maintaining control flow. Algorithm~\ref{alg:seed_mutation} outlines the full similarity-guided block mutation process. A mutation is accepted if the similarity between a mutated block and its original remains above a threshold $T$, allowing the input space to expand. If the similarity does not meet $T$ after multiple attempts, the block is retained without further changes. This approach ensures that mutations remain semantically meaningful while introducing diversity. Block similarity is determined by the similarity of the instructions it contains.

            \begin{algorithm}
            \caption{Similarity-guided Mutation Algorithm}
            \label{alg:seed_mutation}
            \begin{algorithmic}[1]
            \Require Seed corpus $\mathcal{S} = \{S_1, S_2, \ldots, S_K\}$, $T$
            \State $\mathcal{S}^* \leftarrow \emptyset$
            \For{each seed $S \in \mathcal{S}$}
                \State $\{B_1, B_2, \ldots, B_N\} \leftarrow \text{SegmentSeed}(S)$
                \For{$i = 1$ to $N$}
                    \For{$j=1$ to $R$}
                       \State $B_i^{*} \ \leftarrow \textbf{BlockMutation}(B_i) $
                        \State $sim \leftarrow \textbf{BlockSimilarity}(B_i, B_i^{*})$
                        \If{$sim < T$}
                            \State $B_i \leftarrow B_i^{*}, break$
                        \EndIf
                    \EndFor
                \EndFor    
                \State $S' \leftarrow \text{ReassembleSeed}(\{B_1, B_2, \ldots, B_N\})$
                \State $\mathcal{S}^* \leftarrow \mathcal{S}^* \cup \{S'\}$
            \EndFor
            \State \Return $\mathcal{S}^*$
            \end{algorithmic}
            \end{algorithm}

    Instruction-level semantic similarity, modeled across execution units, reflects the similarity of the processor state transitions induced by the instructions. Based on this observation, we design an instruction similarity metric that approximates processor state-space similarity by analyzing instruction semantics at multiple granularities.
    Given two RISC-V instructions, Instruction A ($I_a$) and Instruction B ($I_b$), the similarity function decomposes instruction similarity into four components: instruction type, opcode, sub-semantic, and field-level similarity.
        \begin{equation}
            \begin{split}
                S(I_a, I_b) = &\ w_1 S_{\text{tp}}(I_a, I_b) + w_2 S_{\text{op}}(I_a, I_b) + \\
                &\ w_3 S_{\text{sm}}(I_a, I_b) + w_4 S_{\text{F}}(I_a, I_b)
            \end{split}
            \label{eq:inst_sim}
        \end{equation}

    \noindent\paragraph{Type similarity.} 
    Instruction type similarity measures the similarity between the encoding layouts of two instructions. In RISC-V, instruction types (e.g., R, I, S) define the syntactic structure of instructions, including the number and positions of source registers, destination registers, and immediate fields. Instructions sharing the same type are assigned the highest type similarity, as they follow identical encoding layouts. For instructions of different types, similarity is determined based on the structural overlap of their encoding layouts.

    \noindent\paragraph{Opcode similarity.}
    Opcode similarity captures the coarse-grained functional relatedness between instructions. The opcode primarily defines the high-level operation class of an instruction, such as arithmetic computation, memory access, control flow, or system operations. Instructions with identical opcodes are assigned the highest opcode similarity. Opcodes within the same functional category receive partial similarity. Instructions with unrelated opcodes are assigned low similarity.

    \noindent\paragraph{Sub-semantic similarity.}
    While opcode similarity captures coarse functional behavior, it cannot distinguish fine-grained semantic differences among instructions. To address this, we introduce sub-semantic similarity, which models instruction semantics at a finer granularity by explicitly considering processor execution units. Instructions are first grouped by their primary execution units. Instructions dispatched to the same execution unit are assigned higher sub-semantic similarity, while instructions targeting different units receive lower similarity. Within the same execution unit, sub-semantic similarity further differentiates instructions based on operational characteristics, such as arithmetic operation type, comparison semantics, or data movement patterns.

    \noindent\paragraph{Filed similarity.}
    Finally, field-level similarity measures the similarity of instruction operands and encoding fields (e.g., registers and immediates), capturing variations that may influence data dependencies and microarchitectural behavior without changing the instruction’s core semantics.

    The field similarity is computed using the normalized Hamming distance between the corresponding instruction fields:
    \begin{equation}
        S_{\text{F}}(I_a, I_b) = 
        \sum_{i \in \mathcal{F}} w_i \left( 1 - H(f_{i}^{(a)}, f_{i}^{(b)}) \right)
        \label{eq:sfield}
        \end{equation}
    
    \noindent
    In~(\ref{eq:sfield}):

    \begin{itemize}
        \item $\mathcal{F}$ denotes the set of considered instruction fields, such as \texttt{func3}, \texttt{func7}, \texttt{func2}, and \texttt{operands}.
        
        \item $f_i^{(a)}$ and $f_i^{(b)}$ are the values of field $i$ in instructions $I_a$ and $I_b$, respectively.
        
        \item $H(x, y)$ denotes the normalized Hamming distance between two bit-vectors $x$ and $y$, defined as:
        \[
        H(x, y) = \frac{\text{HammingDist}(x, y)}{\text{bitlength}(x)}
        \]
        
        \item $w_i$ is the weight assigned to field $i$, subject to:
        \[
        \sum_{i \in \mathcal{F}} w_i = 1
        \]
    \end{itemize}

    The overall similarity~(\ref{eq:inst_sim}) is a weighted combination of these measures, capturing both syntactic structure and semantic intent. This fine-grained metric enables effective clustering and mutation of instructions, producing valid and diverse test cases.

\subsection{Differential Testing}
    \label{subsec:difftest}
    After mutation, blocks are reassembled into complete test cases and executed on both RTL processor simulators and ISA reference simulators. Discrepancies in their behavior indicate potential bugs. Using this differential testing framework, SimFuzz is able to reveal inconsistencies in Rocket, BOOM, and XiangShan, uncovering previously unknown bugs.

\section{Evaluation}
 We implemented SimFuzz for the RISC-V GC instruction set, which is supported by all general-purpose processor. The evaluation focuses on two aspects: coverage (Section~\ref{subsec:coverageeva}) and bug discovery (Section~\ref{subsec:bugdiscovery}).

\subsection{Coverage}
\label{subsec:coverageeva}

\subsubsection{Setting}
    Coverage is measured using the xfuzz testing framework~\cite{xfuzz}, which supports multiple coverage metrics: \texttt{line}, \texttt{mux}, \texttt{toggle}, and \texttt{branch}. We focus on mux and toggle coverage, as these metrics are specifically designed for hardware and evaluated by other works.

    The seed corpus includes 42 real-world, historical bug-triggering test cases (Historical) that we collected, along with test cases generated by the random C program generator Csmith~\cite{csmith-repo}, used by the BOOM project, and test cases produced by Cascade~\cite{cascade}, a state-of-the-art processor fuzzing test case generation framework.
    
\subsubsection{Corpus evaluation}
\begin{figure}[tb]
          \centering
          \includegraphics[width=0.48\textwidth]{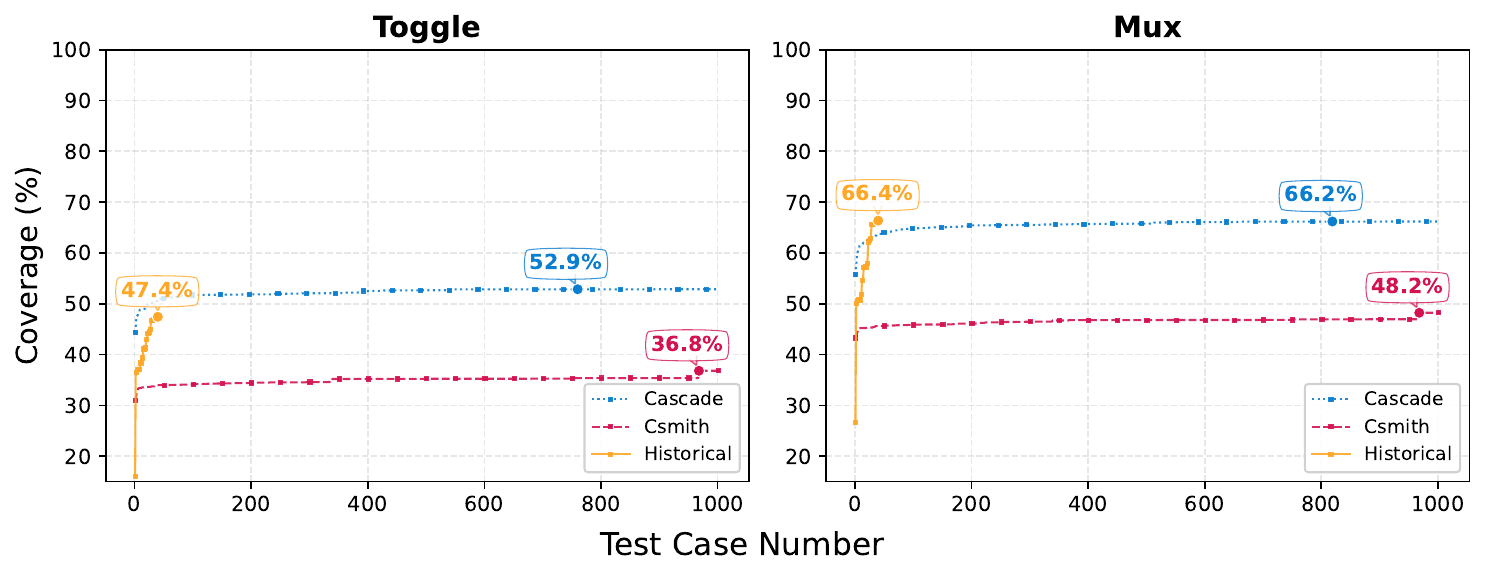}
          \caption{Toggol and mux coverage across different seed corpora without fuzzing. The Historical seed corpus achieve a maximum mux coverage of 66.4\%. }
          \label{fig:seedcorpus}
        \end{figure}

    We use Csmith and Cascade to generate 1,000 seeds each, along with 42 Historical seeds. Fig.~\ref{fig:seedcorpus} shows the total coverage across the three corpora. The Historical seeds achieved a maximum mux coverage of 66.4\%, comparable to Cascade (66.2\%) and significantly higher than Csmith (48.2\%). Despite its smaller size, the Historical seeds produced strong results, highlighting the value of test cases that have previously exposed real bugs for seed construction.
    
\subsubsection{Design evaluation}

    \begin{figure}[hb]
        \centering
        \subfigure[Mux coverage]{
            \includegraphics[width=0.475\textwidth]{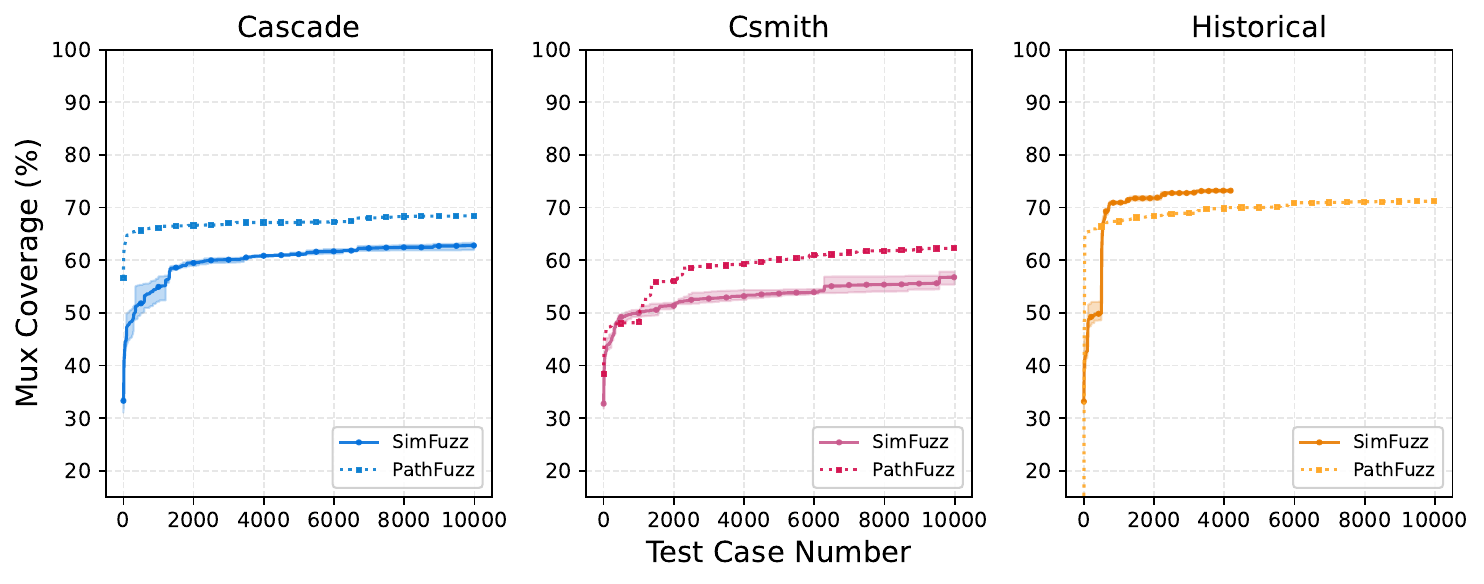}
            \vspace{-0.5cm}
            \label{fig:mux_coverage }}
        \subfigure[Toggle coverage ]{
            \includegraphics[width=0.475\textwidth]{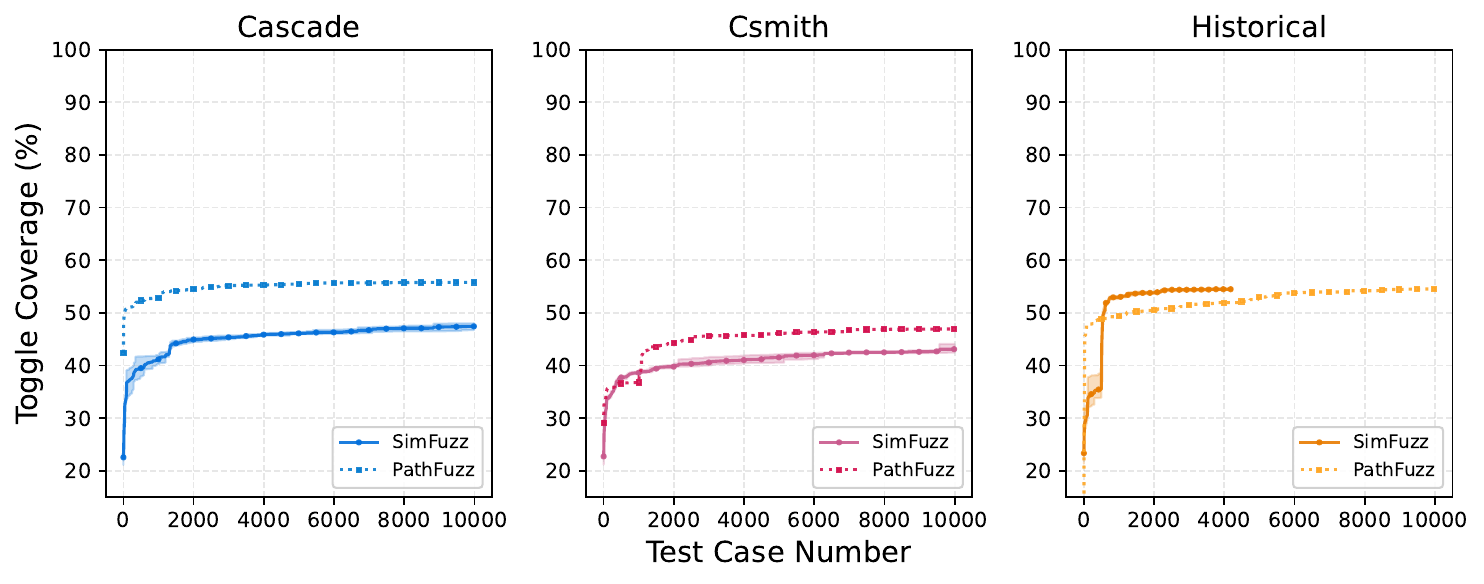}
            \label{fig:toggol_coverage}
        }
        \caption{SimFuzz vs. PathFuzz coverage. (a) Mux: SimFuzz peaks at 73.22\% (historical). (b) Toggle: SimFuzz peaks at 54.51\% (historical).}
        \label{fig:coverages}
    \end{figure}
    We design an input space similarity matrix to guide the mutation process and assess whether direct exploration of the processor input space can effectively cover processor states. In our experiments, we set the similarity threshold to $T=0.5$ based on preliminary tests.
    Csmith and Cascade were used to generate 1,000 seeds each. For SimFuzz, each seed from the Csmith and Cascade corpora was mutated 10 times, producing a total of 10,000 test cases. Historical seeds were mutated 100 times each, resulting in 4,200 test cases.
    For comparison, we evaluate SimFuzz against the state-of-the-art processor fuzzer PathFuzz~\cite{pathfuzz}. PathFuzz first flattens a seed’s memory access footprint into a linear sequence and then applies coverage-guided mutations. We ran PathFuzz for 10,000 executions per seed type.
        
    \begin{table}[t]
    \setlength{\tabcolsep}{5pt}
        \caption{Comparison of SimFuzz with PathFuzz across different seed corpora.}
        \label{tab:fuzzer-comparison}
        \begin{tabular}{cccccc}
        \toprule
        \textbf{Seed Corpus} & \textbf{\makebox[1.2cm]{Coverage}} & \textbf{SimFuzz(\%)} & \textbf{\makebox[1.2cm]{PathFuzz(\%)}} & \textbf{\makebox[1.2cm]{Diff. (\%)}}\\ 
        \multirow{2}{*}{Cascade} 
            & Mux     & 62.80 ±0.69 & 68.44 & -5.64 & \\
            & Toggle  & 47.46 ±0.65 & 55.81 & -8.35 & \\
        \cmidrule(lr){1-5}
        
        \multirow{2}{*}{Csmith}
            & Mux     & 56.78 ±1.24 & 62.31 & -5.53 & \\
            & Toggle  & 43.12 ±0.81 & 46.95 & -3.83 & \\
        \cmidrule(lr){1-5}
        
        \multirow{2}{*}{Historic}
            & Mux     & 73.22 ±0.27 & 71.33 & +1.89 & \\
            & Toggle  & 54.51 ±0.13 & 54.57 & -0.06 &\\
        \bottomrule
        \end{tabular}
        
        \begin{flushleft}
        \small
        \textsuperscript{*}$\pm$ values represent half of the confidence interval range. \\
        \textsuperscript{‡}Historic seeds: SimFuzz (4,200 tests), others (10,000 tests).
        \end{flushleft}
    \end{table}

    Fig.~\ref{fig:coverages} shows the mux and toggle coverage achieved by SimFuzz and the coverage-guided fuzzer PathFuzz after the same number of executions. Unlike PathFuzz, SimFuzz does not rely on coverage feedback, yet it achieves higher coverage on the Historical seeds. Detailed numerical results are presented in Table~\ref{tab:fuzzer-comparison}. These results indicate that SimFuzz achieves coverage comparable to PathFuzz across most corpora and even slightly surpasses it on Historical seeds, demonstrating that similarity-guided exploration can effectively reach diverse processor states without relying on coverage feedback.

\begin{figure}[pb]
          \centering
          \includegraphics[width=0.47\textwidth]{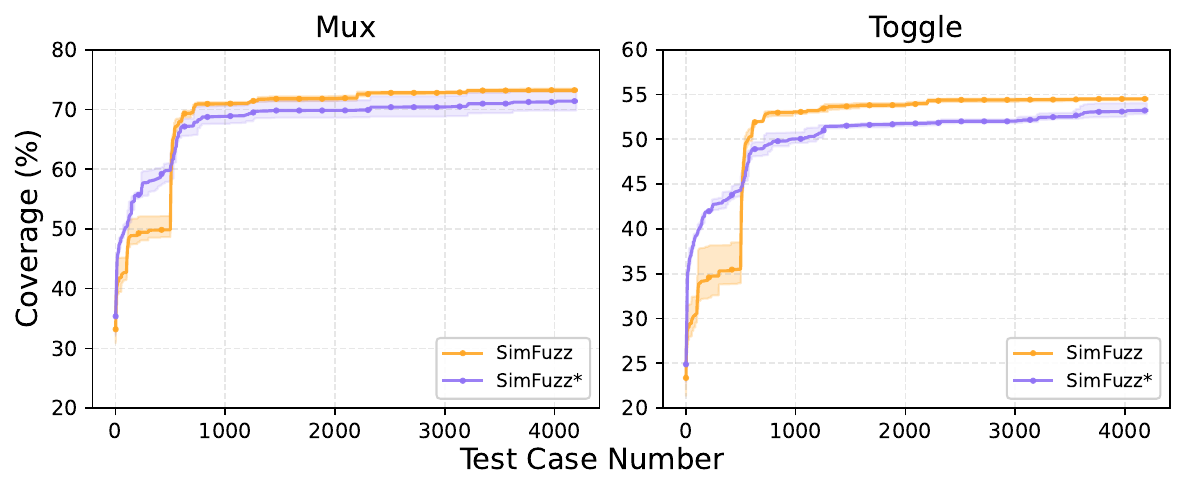}
          \caption{SimFuzz vs. SimFuzz* on mux and toggle coverage. }
          \label{fig:simfuzz*}
        \end{figure}

    To further evaluate the effectiveness of the similarity-guided block-level mutation strategy, we compared SimFuzz with its variant SimFuzz*, in which similarity guidance is disabled. As shown in Fig.~\ref{fig:simfuzz*}, SimFuzz consistently outperforms SimFuzz* on the Historical seeds. However, the overall coverage gap between the two remains moderate. A similar trend is observed for PathFuzz: once coverage reaches a certain level, it gradually saturates with little further improvement.

    This saturation phenomenon can be attributed to two main factors. First, some processor states are inherently difficult to trigger. Second, given the limited input space and mutation strategies, most easily reachable states may already have been explored, while the remaining states often require more complex semantic perturbations or intricate inter-instruction dependencies to be triggered.
    Moreover, this observation suggests that coverage alone may not sufficiently characterize the extent of processor state exploration. A systematic investigation of the validity and effectiveness of coverage metrics for processor fuzzing remains an important direction for future work.

\begin{table}[tbp]
    \centering
    \caption{A list of bugs discovered by SimFuzz. The “From” column indicates the seed origin: S denotes a seed from the constructed corpus, B denotes a bug discovered by SimFuzz.}

    \label{tab:buglist}
    \renewcommand{\arraystretch}{1.2}
    \resizebox{\linewidth}{!}{%
    \begin{tabular}{
        l
        l
        c
        c
        c
        c
    }
    \toprule
    \textbf{Processor} & \textbf{Afftected~Unit} & \textbf{From} & \textbf{New} & \textbf{Confirmed} & \textbf{Fixed}\\
    \midrule
    \multirow{1}{*}{\textbf{BOOM}}  & \textbf{B1:} Floating & S1 & \ding{55} & \ding{55} & \ding{55} \\
    \midrule
    \multirow{4}{*}{\textbf{XiangShan}}
        & \textbf{B7:} Decode   & S5  & \ding{51} & \ding{51} & \ding{51} \\
        & \textbf{B11:} Backend & NA  & \ding{51} & \ding{51} & \ding{51} \\
        & \textbf{B14:} Decode & B14 & \ding{51} & \ding{51} & \ding{51} \\
        & \textbf{B16:} Load & NA  & \ding{51} & \ding{51} & \ding{51} \\
    \midrule
    \multirow{13}{*}{\textbf{NEMU}}
       & \textbf{B2:} Floating & S4 & \ding{55} & \ding{55} & \ding{51} \\
       & \textbf{B3:} Control & S1 & \ding{51} & \ding{51} & \ding{51} \\
       & \textbf{B4:} Control & S6 & \ding{51} & \ding{51} & \ding{51} \\
       & \textbf{B5:} Decode & NA & \ding{55} & \ding{55} & \ding{51} \\
       & \textbf{B6:} Memory & NA & \ding{51} & \ding{51} & \ding{51} \\
       & \textbf{B8:} Decode & S5 & \ding{51} & \ding{51} & \ding{51} \\
       & \textbf{B9:} Control & S7 & \ding{51} & \ding{51} & \ding{51} \\
       & \textbf{B10:} Control & B10 & \ding{51} & \ding{51} & \ding{51} \\
       & \textbf{B12:} Decode & NA & \ding{51} & \ding{51} & \ding{51} \\
       & \textbf{B13:} Decode & B11 & \ding{51} & \ding{51} & \ding{51} \\
       & \textbf{B15:} Decode & B14 & \ding{51} & \ding{51} & \ding{51} \\
       & \textbf{B17:} Memory & S8 & \ding{51} & \ding{51} & \ding{51} \\
    \bottomrule
    \end{tabular}
    }
\end{table}

\subsection{Bug Discovery}
\label{subsec:bugdiscovery}
    
    SimFuzz discovers a total of 17 bugs on real processors and related reference models, including 3 previously known bugs and 14 newly identified ones. All newly found bugs have been reported to the community, confirmed, and fixed. 

    The seed corpus consist of 42 binary test cases from 37 bug reports of three open-source processors.
    Rocket~\cite{rocketrepo} and BOOM~\cite{boomrepo} are early open-source RISC-V processors with over a decade of development, while XiangShan~\cite{XiangShanWebsite} is a more recent processor introduced within the last five years. These three processors are commonly used as targets in verification studies~\cite{hur2021difuzzrtl,morfuzz,pathfuzz}. Rocket and BOOM rely on the Spike~\cite{SpikeISASim} simulator as a reference model, whereas XiangShan uses NEMU~\cite{nemurepo}. The correctness of the reference model is critical, as errors in the model can mask real processor bugs. Notably, some bugs are observed in both XiangShan and NEMU, emphasizing the importance of reference model accuracy.
    Table~\ref{tab:buglist} summarizes all identified bugs. Since all bugs are discovered prior to processor fabrication, no exploits currently exist that can be executed in real-world scenarios under normal user privileges. The following section provides detailed descriptions of several RTL-level bugs.

\label{subsec:xiangshanbugs}

\noindent\textbf{Bug B7, B8.} These two bugs involve the AES encryption instruction \texttt{aes64ks1i}~\cite{manual_unprivileged} in the XiangShan processor and NEMU simulator. The \texttt{rnum} field specifies the substitution round of the SBox~\cite{manual_unprivileged} and is 5 bits wide (range \texttt{0-31}), but the specification allows only \texttt{0-0xA}. Both XiangShan and NEMU fail to enforce this restriction, permitting out-of-range values (e.g., \texttt{rnum = 0x15}) to execute normally. Notably, this bug is previously observed in Rocket and has reappeared in XiangShan, highlighting the difficulty of detecting certain corner cases in the RISC-V specification. SimFuzz leverages cross-processor bug data to cover such corner cases effectively.

\noindent\textbf{Bug B11.} This bug originates from a bug in XiangShan’s back-end control logic. When a \texttt{jalr} instruction performs an indirect jump to an invalid physical address, the processor fails to trigger exception handling correctly. Specifically, if the computed target address exceeds the 48-bit physical address range, an instruction access fault should occur. However, the back-end control logic does not treat this as a valid pipeline redirection event. The valid signal generation only accounts for branch mispredictions, preventing timely pipeline flushing and proper exception handling. 

\noindent\textbf{Bug B13, B14.} According to the RISC-V privileged specification~\cite{manual_privileged}, the indirect register selectors \texttt{siselect} and \texttt{vsiselect} in S-mode and VS-mode should allow addresses \texttt{0x0-0xFFF}. In XiangShan, hardware restricts this range to \texttt{0x0-0x1FF}, while NEMU imposes no limit. This discrepancy renders critical features, such as the advanced interrupt architecture, unusable on the processor, representing a severe implementation bug.

\noindent\textbf{Bug B16.} A bug in XiangShan’s Load Unit causes incorrect values in the \texttt{mtval} register during exceptions. When a load instruction (e.g., \texttt{fld}) triggers both an address misalignment and a page table translation miss, \texttt{mtval} records an unrelated address rather than the original faulting memory access. Since the OS exception handler relies on \texttt{mtval} to handle page faults or permission violations, this error can cause incorrect fault handling and potential failure to service user-level programs.

\subsection{Other Findings}
 
    SimFuzz revealed that processors from different open-source projects contain bugs stemming from the same underlying issues. Moreover, flaws related to known bugs persisting even in newer versions. This finding suggests that under the same ISA, different processors may share identical or similar bugs. Therefore, during processor design and verification, particular attention should be paid to known bugs to prevent the recurrence of similar errors. Expanding such corpora with additional historical bug-triggering seeds could further improve coverage and reduce verification time. We plan to enlarge the seed corpus by collecting seeds from more processors to enhance processor security and further reduce verification time.

\section{Conclusion}
    We present SimFuzz, a novel processor fuzzing approach. Unlike traditional coverage-guided strategies, SimFuzz constructs a high-quality seed corpus using historical bug-triggering test cases from real-world processors. It then employs a similarity metric to explore the input space while preserving the control-flow structure of these seeds during mutation, enabling deeper coverage of processor states.
    Experimental results show that SimFuzz achieves higher coverage than coverage-guided methods, discovers 14 previously unknown bugs, and obtains 7 CVE identifiers, demonstrating both its effectiveness and practical significance. Furthermore, our results indicate that identical or similar bugs can arise across processors sharing the same architecture, highlighting the importance of incorporating real-world bug samples into processor verification.

\section{Acknowledgment}
    The authors would like to thank all anonymous reviewers for their valuable comments and suggestions, and the maintainers of OpenXiangShan community for their support in understanding and fixing bugs. Additionally, we acknowledge all open-source works (including but not limited to~\cite{Kevin18rfuzz, hur2021difuzzrtl, cascade, pathfuzz, morfuzz}) related to processor fuzzing. This paper is supported by the National Key R\&D Program of China (2024YFB4506200) and the YuanTu Large Research Infrastructure.

\bibliographystyle{plain}
\bibliography{reference}

\end{document}